\documentstyle[a4,epsf]{article}  
\begin{document}                
\newcommand{\manual}{rm}        
\newcommand\bs{\char '134 }     
\newcommand{\Het}{$^3{\mathrm{He}}$}
\newcommand{\Hef}{$^4{\mathrm{He}}$}
\newcommand{\A}{{\mathrm{A}}}
\newcommand{\D}{{\mathrm{D}}}
\newcommand{\simlt}{\stackrel{<}{{}_\sim}}
\newcommand{\simgt}{\stackrel{>}{{}_\sim}}
\renewcommand{\floatpagefraction}{1.}
\renewcommand{\topfraction}{1.}
\renewcommand{\bottomfraction}{1.}
\renewcommand{\textfraction}{0.}               
\renewcommand{\thefootnote}{{\mathrm F}\arabic{footnote}}       
\title{Effects of a dynamical role for exchanged quarks and nuclear
gluons in nuclei: multinucleon correlations in deep-inelastic lepton
scattering}
\author{Saul Barshay and Georg Kreyerhoff\\
III. Physikalisches Institut\\
RWTH Aachen\\
D-52056 Aachen\\
Germany}
%
\maketitle
\begin{abstract}                
 It is shown that new data from the HERMES collaboration, as well
as all of the earlier improved data from experiments concerning
the EMC effect and shadowing in deep-inelastic scattering of leptons
from nuclei, provide strong evidence for an explicit dynamical
role played by exchanged quarks and nuclear gluons in the basic, tightly-bound
systems of three and four nucleons, $^3{\mathrm{He}}$ and $^4{\mathrm{He}}$.
This opens the way for specific quark-gluon dynamics instigating multinucleon
correlations in nuclei.
\end{abstract}
New experimental data have appeared \cite{ref1}, relating to the unusual
behavior of nucleons and their constituents in atomic nuclei. The
unusual behavior appears when nuclei are probed by deep-inelastic
scattering of charged leptons\cite{ref2,ref3,ref4,ref5,ref6}. The
original discovery\cite{ref2}, called the ``EMC effect'', is a depression
below unity of the ratio of cross sections (per nucleon) 
$\frac{\sigma_\A}{\sigma_\D}$, for nucleus A as compared to deuterium, D.
This occurs\cite{ref2,ref3} in the domain of momentum-fraction $x$,
which is characteristic of valence quarks in a free nucleon, $x\simgt 0.25$.
A depression in $\frac{\sigma_\A}{\sigma_\D}$ also occurs\cite{ref4,ref5}
in the domain of small $x\simlt 0.06$, at low and moderate $Q^2$,
the negative of the squared four-momentum transfer from the lepton.
This effect is usually referred to as ``shadowing''. An effect of this
kind is expected at small $x$, at and near to $Q^2=0$, as a geometrical
effect of the nuclear surface, which diminishes the intensity
of vector-meson-like components of the photon through strong interactions
with nucleons near to this surface
\footnote{This particular effect, accomodated by conventional concepts in
nuclear physics, is not dealt with in this paper, except for an observation
at the end concerning a possible difference between ``shadowing'' for real
photons $(Q^2=0, \vec{Q}\ne 0)$ and for virtual photons with $Q^2$ near zero ($\vec{Q}\to 0$).
}. From the very 
beginning\cite{ref2,ref3} of the experimental measurements, there have
been two important observations (which often have not been emphasized
in subsequent work):
\begin{itemize}
\item[(1)] The ``EMC effect'' is conspicuously prominent\cite{ref3,ref4}
in the basic nucleus \Hef; its growth with A in going to $^{40}$Ca and
to $^{119}$Sn is moderate. A corollary to this is the matter
of possible dependence upon $Q^2$ of the loss of momentum fraction
from valence quarks in nuclei. There have been indications in the 
data\cite{ref8}, that a small additional loss occurs as $Q^2$ is increased,
that is $\frac{\sigma_\A}{\sigma_\D}$ is depressed further below unity.
This effect, although small, is ``important because of its direction:
if the loss of valence-quark momentum fraction \underline{increases}
with $Q^2$, the underlying dynamics can hardly be solely
conventional nuclear physics''.\cite{ref8}
\item[(2)]There have been indications in data\cite{ref9,ref10} of a rather
sharp and strong drop of $\frac{\sigma_\A}{\sigma_\D}$ below unity,
in the domain $0.01<x<0.1$ for $0.05<Q^2<1.5 (\frac{\mathrm{GeV}}{c})^2$.
This suggests the possibility that an important part of the ``shadowing''
effect has an origin in specifically partonic dynamics within nuclei,
an origin that should be a natural extension of the partonic dynamics
which leads to the loss of momentum fraction in the $x$-domain of
the valence quarks\cite{ref8}. 
\end{itemize}
The above facts point to the possibility that quark-gluon degrees of 
freedom contribute explicitly to multinucleon correlations\cite{ref11}
and forces\cite{ref12} in nuclei. These correlations\cite{ref11}
and forces are prominently present already in the basic, tightly-bound
He nuclei. When one recalls that the maximum nucleon densities
encountered in $^3$He and $^4$He are about two times higher than those
in any other nucleus, it is not surprising that if relatively short-range,
three-body correlations of quark-gluonic origin exist, there are significant
effects observable in these systems, in particular when probed by 
deep-inelastic lepton scattering. This was the basis for a detailed
phenomenological model\cite{ref11} for a three-nucleon correlation 
(force\cite{ref12}) which involved as dynamics the exchange of one
quark from each nucleon, under the influence of nuclear gluon interactions
\footnote{A diagrammatic representation is in Fig.~2 of Ref.~7 and
Fig.~2 of Ref.~10}. 
Momentum fraction is lost from charged constituents to
the quanta of a nuclear gluonic field.\par
The dynamical model offers the possibility of understanding the essential
EMC effect at $x>0.25$, and also a part of the shadowing-like
effect, in a \underline{unified} picture\cite{ref8}. A quantitative 
description (on a $\chi^2$ basis) of all of the nuclear data in the domain
$0.02 \le x \le 0.7$ was achieved with a few parameters whose values are 
estimated {\it a priori} from physical considerations, 
and whose fit values agree with these estimates\cite{ref11}. 
The A-dependence of $\frac{\sigma_\A}{\sigma_\D}$
is successfully predicted, as has been shown by the detailed analyses of data 
by Smirnov\cite{ref14}. New predictions are made\cite{ref11}. One,
which has been experimentally verified\cite{ref15,ref8}, involves
the close similarity between $\frac{\sigma_\A}{\sigma_\D}$ for the 
``irregular'' nucleus $^6$Li and that for $^4$He. The nucleus $^6$Li
is irregular in having an r.~m.~s.~radius of $\sim 2.5$ fm, about the
same as that for $^{12}$C. Thus, \underline{if} viewed as a system of 
uniform nucleon density, this density is low, about one-half
that of $^4$He. From this point of view, one would expect little
reduction in $\frac{\sigma_\A}{\sigma_\D}$ for $^6$Li. Instead the data
show that on ``a bin to bin comparison the $\frac{\mathrm {Li}}{\mathrm{He}}$
ratio is consistent with unity over the common $x$ range''\cite{ref15}
$(0.01\simlt x\simlt 0.5)$. (Note Fig.~1 of Ref. 7.) Thus, as probed
by deep-inelastic lepton scattering, $^6$Li behaves as if composed
of a $^4$He-like group (and/or $^3$He-like groups) of nucleons.\par
Quantitative examples of the increased loss of momentum fraction from
valence quarks (reaching a limit) with increasing $Q^2$, were given
in Figs. 7, 8 of Ref.~10, for $^4$He and $^{12}$C in particular. The new
data from HERMES\cite{ref1} give the most distinct indication of this
effect, for $^3$He and $^{14}$N, over the range 
$1<Q^2<20(\frac{\mathrm{GeV}}{c})^2$, for $\langle x \rangle \sim 0.4$.
We describe this below. The new data from HERMES\cite{ref1} exhibit a 
sharp and strong drop of $\frac{\sigma_\A}{\sigma_\D}$ below unity
for $x<0.06$, The data show a marked $Q^2$ dependence of this effect for 
(average) $x$ values in the domain $0.01 < \langle x \rangle < 0.06$.
What is strikingly unusual is that the depression of 
$\frac{\sigma_\A}{\sigma_\D}$ in this domain of $\langle x\rangle$,
\underline{first increases} as $Q^2$ moves up from near zero,
$0.3 < Q^2 < 1.5(\frac{\mathrm{GeV}}{c})^2$. A principal purpose
of this paper is to describe this effect quantitatively, and to
discuss physically two possible, complementary origins of such 
unusual $Q^2$ dependence.
This is done below, within the unified framework provided by the model for 
three-nucleon correlations which originate in explicit quark-gluon
dynamics, in $^3$He in particular. The new HERMES data\cite{ref1} show
that $\frac{\sigma_\A}{\sigma_D}$ already falls significantly below
unity in $^3$He, as in $^4$He, which was anticipated\cite{ref11,ref14}.\par
To start the analysis, we briefly explain the essential physical
features of the formula\footnote
{In the comparision with data carried out in Ref.~10, this
formula is identified with the ratio of structure functions
$\frac{F_2^\A(x,Q^2)}{F_2^\D(x,Q^2)}$. This is identical with $\frac{\sigma_\A}{\sigma_\D}$ if the
ratio of longitudinal to transverse deep-inelastic scattering cross sections
$\tilde{R}^\A = \frac{\sigma^\A_{\mathrm{L}}}{\sigma^\A_{\mathrm{T}}}$ satisfies
$\tilde{R}^\A(x,Q^2)=\tilde{R}^\D(x,Q^2)$ (or independently of such an equality,
if $Q^2\to 0$). Here, we use the phenomenological formula for $\frac{\sigma_\A}
{\sigma_\D}$, and we explicitly discuss the matter (Ref.~1) of $\tilde{R}^\A(x,Q^2)$ possibly
not equal to $\tilde{R}^\D(x,Q^2)$ later in this paper. 
}
for $\frac{\sigma_\A}{\sigma_\D}$
which provides a quantitative description of all of the nuclear data
in the domain $0.02 \le x \le 0.7$, for 
$1 < Q^2 < 100 (\frac{\mathrm{GeV}}{c})^2$.
\begin{eqnarray}
\frac{\sigma_\A(x, Q^2)}{\sigma_\D(x, Q^2)} = 1&-&\delta(A)\left\{
\frac{3.3\sqrt{x}(1-x)^3-C(Q^2)f(Q^2,x^2)\sqrt{x}{\mathrm{e}}^{-Bx^2}}
{3.3\sqrt{x}(1-x)^3+1.1(1-x)^7}\right\}\nonumber\\
&-&\delta(A)\left\{ \frac{1.1(1-x)^7-\tilde{C} x^\beta {\mathrm{e}}^{-B'x^2}}
{3.3\sqrt{x}(1-x)^3+1.1(1-x)^7}\right\}
\end{eqnarray}
with $B=11.3$, $B'=35$, $\beta=0.35$, 
\begin{displaymath}
\delta(A) =  0.27\left\{ 1-\frac{1}{A^{1/3}}-\frac{1.145}{A^{2/3}}+
\frac{0.93}{A}+\frac{0.88}{A^{4/3}}-\frac{0.59}{A^{5/3}}\right\}
\end{displaymath}
and
\begin{eqnarray*}
C(Q^2) &=& \frac{3}{\int_0^1 \frac{dx}{x} \sqrt{x} {\mathrm{e}}^{-11.3x^2}
f(Q^2,x^2) } \\
f(Q^2,x^2)&=& {\mathrm{e}}^{-48x^4\left(\frac{\ln(Q^2/2)}{\ln(Q^2/0.04)}\right)}
\\
\tilde{C}&\cong& \frac{1.1\int_{0.06}^1 \frac{dx}{x} (1-x)^7}{\int_{0.06}^1
\frac{dx}{x} x^{0.35} {\mathrm{e}}^{-35x^2}}
\end{eqnarray*}
In this formula for $\frac{\sigma_\A}{\sigma_\D}$, the second term modifies
the usual valence-quark distribution \cite{ref17} of momentum fraction. While
maintaining the valence-quark number through $C(Q^2)$, a fraction
$\delta(A)$ of the valence quarks (per nucleon) are removed
from the usual distribution for a free nucleon, and are distributed
instead in a Gaussian form\footnote{The approximation of using a Gaussian form in momentum space is based upon
an approximate Gaussian form for quark motion in configuration space in the three-nucleon
correlation. The Gaussian form is not zero at $x=1$, but the contribution of this
term to $\sigma_\A$ is then $\propto {\mathrm {e}}^{-11.3}$, which is negligible
compared to the non-zero contribution at $x=1$ due to the nuclear Fermi motion. The
Gaussian form does simulate the effect of Fermi motion in the ratio $\frac{\sigma_\A}{\sigma_\D}$,
causing it to move sharply upward above unity as $x\to 1$. 
}
characterized by a size parameter
$B$; the size characterizes the spatial motion of a fraction of the quarks
in a correlated, three-nucleon system. The fit parameter corresponds
to a spatial dimension $(\cong \sqrt{3B}/m_N)$ of about 1.2 fm, the
same as that estimated {\it a priori} from geometrical considerations
concerning a three-nucleon correlation\cite{ref11}, (and is well
within the r.~m.~s.~charge radius of $^3$He, which is about 1.7 fm). The
modification results in a gradual depression of $\frac{\sigma_\A}{\sigma_\D}$
below unity in the valence-quark domain, $x\simgt 0.3$. This depression
increases a little with increasing $Q^2$, in particular in the domain
$1<Q^2<20 (\frac{\mathrm{GeV}}{c})^2$, reaching a limiting depression
from above, as is illustrated in Figs.~7, 8 of Ref.~10 for $^4$He and
$^{12}$C, respectively. The effect, which is due to $f(Q^2,x^2)$ in
the second term in Eq.~(1), arises physically because the correlation
among valence quarks and nuclear gluons, as seen by a high-$Q^2$ probe,
can be considered to be an additional nuclear degree of freedom, whose
$x$-distribution ``softens'' as $Q^2$ increases due to more momentum
fraction being lost to nuclear gluons. In $f(Q^2,x^2)$, the loss is
limited by the decreasing strength of the gluon coupling, $\alpha_s(Q^2)
\propto \frac{1}{\ln(Q^2/0.04)}$ (for $\Lambda_{\mathrm{QCD}}\sim 200
\frac{\mathrm{MeV}}{c}$)\footnote{ $f(Q^2,x^2)=1$ for $Q^2=2(\frac{\mathrm{GeV}}{c})^2$, and is taken as 1 for
 $Q^2$ below this value in the present analysis of the new data down to $Q^2\sim 0.3(\frac{\mathrm{GeV}}{c})^2$.}.
The A dependence is given by
the function $\delta(A)$, which uniquely arises by excluding from
the three-nucleon correlation the number of nucleons ${\mathrm{A}}_S$,
which reside in the (relatively diffuse) surface of a large nucleus,
that is\cite{ref11} $\delta(A)=N(1-{\frac{{\mathrm A}_S}{\mathrm{A}}})$.
The single, overall normalization parameter, $N=(27\pm 6)\%$, is
fit\cite{ref11} to $\delta({\mathrm{A}}=4) = 6^{+1.5}_{-1}\%$. (Then,
$\delta({\mathrm{A}}=14)\cong 12.6 \%$ and $\delta({\mathrm{A}}=3)
\cong 4.2\%$.) Let us look now at the HERMES data\cite{ref1}, shown
in Figs.~1,2 for $^{14}$N and $^3$He, respectively. For $\langle x
\rangle \sim 0.4$, well into the domain of valence-quark momentum
fraction, the data indicate that $\frac{\sigma_{\A=14}}{\sigma_\D}$
falls up to $4 \%$ more below unity as $Q^2$ goes up from $\sim 1$
to $\sim 20 (\frac{\mathrm{GeV}}{c})^2$. For $^3$He, the additional
drop may be up
to $2 \%$.\footnote{ These are small changes which might be ignored\cite{ref1}, within the experimental
uncertainties. Nevertheless, the direction is in accord with earlier experimental
indications\cite{ref8}, and the changes are expected within a model of explicit
quark-gluon correlations in nuclei. Note also the $Q^2$ dependence of the $^{119}$Sn data
for $0.4<x<0.5$, in Fig.~4 of M.~Arneodo et.~al.~(NMC), Nucl.~Phys.~{\bf B481}, 23 (1996). }
These numbers appear similar to those calculated
from Eq.~(1) ten years ago\cite{ref11}, in the Figs.~8, 7 for $^{12}$C
and $^4$He, respectively.\par
The physical picture of a three-nucleon correlation of quark-gluonic
origin allows for the natural prediction of a shadowing-like
effect, in fact a rather sharp drop of $\frac{\sigma_\A}{\sigma_\D}$
below unity, for $x<0.06$.\cite{ref11} This is contained in the last
term in Eq.~(1). A colored agglomeration of sea quarks (antiquarks)
is exchanged from each nucleon, instead of a valence quark, under
the influence of nuclear gluon interactions. The fraction of the sea
(taken as $\delta(A)$ in Eq.~(1)) involved in this dynamics is removed
from the relevant momentum-fraction distribution \cite{ref17} for a 
free nucleon and is distributed instead in a form given by the 
$x$-dependence multiplying $\tilde{C}$ in Eq.~(1). The crucial point
is the non-zero power of $x$, $\beta > 0$. Physically, this means
that whereas the number of sea quarks and antiquarks in an individual
nucleon is formally infinite (i.~e.~$\propto \int_0^1\frac{dx}{x}\ldots$),
the number of these which are involved in the three-nucleon correlation
is finite. In the representation for $\frac{\sigma_\A}{\sigma_\D}$ in
Eq.~(1), it is the factor $x^\beta$ which results in the relatively
sharp drop below unity for $x<0.06$. \footnote{The fit parameter $\beta$ is restricted by $0<\beta<\frac{1}{2}$, because
the distribution of momentum fraction in the sea is ``softer'' than that for the valence quarks
($\propto \sqrt{x}$), as $x\to 0$. The initial sea is taken for $x\ge 0.06$. }
The factor 
$\tilde{C}(Q^2)$ ensures that the number of sea quarks and antiquarks
involved in the larger spatial domain of the three-nucleon correlation
is equal to the number removed from the individual nucleon distribution.
The ``size'' parameter $B'$ in the Gaussian, characterizing the spatial
motion of the sea, is expected to be larger than $B$; the fit number
corresponds to a dimension of about $2.2$ fm ($B'$ may increase\cite{ref11}
somewhat with $A$.) Eq.~(1) resulted\cite{ref11} in a representation
with $\chi^2\simlt 1$ per degree of freedom, for all of the nuclear
data taken with $Q^2$ between about 2 and 100 $(\frac{\mathrm{GeV}}{c})^2$,
for $x$ in the domain $0.02 \le x \le 0.7$.\footnote{At smaller $x$, one encounters the need to deal explicitly with the
geometrical shadowing effect which is present already for real photons ($Q^2= 0$, $\vec{Q}\ne 0$).$^{{\mathrm F}1}$}
\par
Now the HERMES data\cite{ref1} exhibit a new $Q^2$ dependence for
$0.3 < Q^2 < 1.5(\frac{\mathrm{GeV}}{c})^2$, with $x$ in the 
domain $0.0126 \le x\le 0.055$, as shown by the first 6 graphs in
Figs.~1,2 for $^{14}$N and $^3$He, respectively. When averaged
over the relevant $Q^2$, the fall of $\frac{\sigma_\A}{\sigma_\D}$ below
unity is even sharper and deeper\cite{ref1} than in most of the earlier
data\cite{ref4,ref5}. We show here that the unusual behavior as a function
of $Q^2$, at small $x$, can be well represented by a simple, overall
form factor, and a normalization change, multiplying the last term in Eq.~(1).
The modification is
\begin{equation}
\delta(A)\times 1 \to 6.3\delta(A)F(Q^2)
\end{equation}
with
\begin{displaymath}
F(Q^2)=\frac{4Q^2}{Q^2+1}\frac{1}{Q^2+1} = 4\left\{1-\frac{1}{Q^2+1}\right\}\frac{1}{Q^2+1}, \;\;\; Q^2\;{\mathrm{in}}\;\left(\frac{\mathrm{GeV}}{c}\right)^2
\end{displaymath}
Keeping all parameters\cite{ref11} fixed, the curves which result from 
Eqs.~(1,2) are superimposed on 
data in Figs.~1,2. Significant aspects of the data both in the valence-quark $x$ domain and
in the shadowing domain are evident in the curves. At the lowest
values of $\langle x \rangle$, the curves are meant to refer specifically
to the new HERMES data\cite{ref1}. The NMC\cite{ref4} and E665\cite{ref5} data
involve a much higher lepton-beam energy, whereas possible explicit dependence
upon this energy is suppressed in the phenomenological Eqs.~(1,2). The data
at these higher energies and at low $Q^2$, may be expected to be closer
to unity because the kinematic variable $\epsilon(x,Q^2,E)$ which
appears in the ratio of cross sections (as discussed in the next
paragraph), is close to unity, thus removing sensitivity$^{{\mathrm F}3}$
to $\tilde{R}^\A(x,Q^2)$. A possible
physical reason for the presence of a form factor which vanishes not only as $Q^2\to\infty$,
but also\cite{ref12} as $Q^2\to 0$ (this is the unusual aspect in Eq.~(2)), lies
in a high degree of coherence associated with the system of exchanged sea partons, and
nuclear gluons (which are individually, colored quanta). This occurs when the system at small
values of $x$ responds to a low-$Q^2$ probe. Qualitatively, exchange of colored quanta
(or aggregates) initially at small $x$, involve spatial dimensions of the order of $\frac{1}{m_Nx}=$
many fermis. Thus coherence over a nucleus, and consequently a form factor falling at least
like approximately $1/Q^2$, may be expected. However, in addition there is the 
\underline{confinement axiom}. Starting with a system of nucleons, individual colored entities
cannot be present over extended dimensions within this system. This suggests that a probe
with $Q^2=0$ ($\vec{Q}=0$) ``sees'' no color-induced correlation. Then, the remaining shadowing is related
to colorless aggregates of quarks which constitute the low-mass vector meson components
of the photon. Such behavior is incorporated phenomenologically in a simple way in Eq.~(2); with $F(Q^2)$
normalized to unity at its maximum, which occurs at $Q^2=1 (\frac{\mathrm{GeV}}{c})^2$ for
this form. Note that averaging $6.3\delta(A)F(Q^2)$ over $Q^2$ in the interval $0.3 < Q^2
< 1.5 (\frac{\mathrm{GeV}}{c})^2$, gives an effective, A-dependent coefficient
of the last term in Eq.~(1) of $\sim 6.1\delta(A)$ (instead of $\delta(A)$); this accounts
quantitatively for the increased depth of the sharp drop for $x<0.06$ in the HERMES
data\cite{ref1} (their Fig.~1) as compared to the earlier data\cite{ref4,ref5} 
\footnote{
 Averaging $6.3\delta(A)F(Q^2)$ over $Q^2$ in the interval
 $ 1 < Q^2< 100 (\frac{\mathrm{GeV}}{c})^2$
 gives an effective, A-dependent coefficient of the last term in Eq.~(1) of $\sim \delta(A)$,
 in agreement with the representation of the earlier data\cite{ref11}. It is worth
 noting that the fit normalization parameter $N\sim 0.3$ in the factor $\delta(A)$
 which multiplies the valence-quark term in Eq.~(1), contains a factor of $\frac{1}{3}$
 corresponding to the {\it a priori} probability for involving a single valence
 quark from each nucleon in the correlation (at any instant). This constraint does not hold
 for the sea quark normalization as changed in Eq.~(3); the fit normalization here is of
 order unity.}.
At
$x\cong 0.01$ the earlier depression of $\frac{\sigma_\A}{\sigma_\D}$ below unity
by about $6\%$ for $^{12}$C is increased to about $35\%$ for $^{14}$N in the HERMES data\cite{ref1}!\par
Within the physical picture embodied in Eq.~(1), we consider the interpretation of the new
data in terms$^{{\mathrm F}3}$ of $\tilde{R}^\A(x,Q^2) > \tilde{R}^\D(x,Q^2)$, as given in the
analysis\cite{ref1} by the HERMES collaboration. The ratio $\frac{\sigma_\A}{\sigma_\D}$
is written in terms of the ratio of structure functions $F_2$, times a multiplicative
factor
\begin{equation}
\frac{\sigma_\A(x,Q^2)}{\sigma_\D(x,Q^2)} = \frac{F_2^\A(x,Q^2)}{F_2^\D(x,Q^2)}\left\{
\frac{(1+\epsilon \tilde{R}^\A(x,Q^2))(1+\tilde{R}^\D(x,Q^2))}{
(1+\epsilon \tilde{R}^\D(x,Q^2))(1+\tilde{R}^\A(x,Q^2))}\right\}
\end{equation}
where $\epsilon(x,Q^2,E)$ is the kinematic variable related to the virtual-photon
polarization, defined by HERMES in their Eq.~(2), \cite{ref1} with $0\le \epsilon\le1$;
($E$ is the initial-lepton laboratory energy). The factor multiplying $\frac{F_2^\A}{F_2^\D}$
is unity independently of the value of $(\tilde{R}^\A-\tilde{R}^\D)$, when the kinematic
variable $\epsilon$ becomes unity; this is formally at $Q^2=0$. As $\epsilon$
moves towards zero, which occurs for larger $Q^2$ in the HERMES data, the factor
becomes less than unity if $\tilde{R}^\A(x,Q^2)>\tilde{R}^\D(x,Q^2)$. However,
if   $(\tilde{R}^\A-\tilde{R}^\D)$ becomes effectively zero already at moderate values of
$Q^2$, then the multiplicative factor is again unity, independently of the value
of $\epsilon$. \footnote{This is in fact, the situation for the values of
$\frac{\tilde{R}^\A(x,Q^2)}{\tilde{R}^\D(x,Q^2)}$ extracted by HERMES under certain assumptions concerning the
behavior of  $\frac{F_2^\A(x,Q^2)}{F_2^\D(x,Q^2)}$ and
$\frac{\tilde{R}^\A(x,Q^2)}{\tilde{R}^\D(x,Q^2)}$. \cite{ref1} Thus, a strong variation
at low $Q^2$ which is evident in the HERMES data in Figs.~1,2, then occurs in
$\frac{\tilde{R}^\A}{\tilde{R}^\D}$.} Assuming that an interpretation of the new data in terms
of  $(\tilde{R}^\A-\tilde{R}^\D)>0$ is valid, we make an approximate 
identification of Eq.~(3)
with the successful phenomenological formula in Eq.~(1), modified as in Eq.~(2).
Then, for $\tilde{R}^\D$ and $\tilde{R}^\A$ less than unity, we obtain an
approximate relation in the domain $x<0.06$, with $Q^2<1.5(\frac{\mathrm{GeV}}{c})^2$, for
a function \footnote{The more complete form of the right-hand side of Eq.~(4) resulting from
Eqs.~(1,2) has two zeros: one at $Q^2\cong 0.043(\frac{\mathrm{GeV}}{c})^2$ which corresponds
to $\epsilon\to 1$; the other at $Q^2\cong 23 (\frac{\mathrm{GeV}}{c})^2$ which corresponds
to $\tilde{R}^\A(x,Q^2)\to \tilde{R}^\D(x,Q^2)$.} $\Delta(x,Q^2)$, at given $E$
\begin{equation}
\Delta(x,Q^2)=\left\{1-\epsilon\right\} (\tilde{R}^\A-\tilde{R}^\D) \sim 6.3\delta(A)s(x)\left\{
1-\frac{1}{Q^2+1}\right\}\left(\frac{1}{Q^2+1}\right)
\end{equation}
where
\begin{displaymath}
s(x)=\frac{1.1(1-x)^7 - \tilde{C}x^{0.35}\mathrm{e}^{-35x^2}}{3.3\sqrt{x}(1-x)^3+1.1(1-x)^7}
\end{displaymath}
The quantity on the right in $\{\ldots\}$ has correct limiting behaviors to phenomenologically,
approximately  represent $\{1-\epsilon\}$ i.~e.~$\epsilon=1$ at $Q^2=0$, and $\epsilon\to 0$
for large $Q^2$ (at given $E$ and $\langle x\rangle$). 
The explicit dependence upon $E$ and $\langle x \rangle$ is suppressed in this approximation;
the numerical values of $\epsilon$ are roughly given as in the HERMES data for the
lowest $\langle x\rangle$-bins\cite{ref1}. 
At fixed small $x$, Eq.~(4) suggests for the ratio
\begin{equation}
\left\{ \frac{\tilde{R}^\A(x,Q^2)}{\tilde{R}^\D(x,Q^2)}-1\right\} \propto \delta(A)\left(
\frac{1}{Q^2+1}\right)
\end{equation}
Eq.~(5) is an approximate representation$^{\mathrm{F}11}$ of the growth at low $Q^2$ for 
the ratio at small $x$,
which is extracted from the $Q^2$ variation of $\frac{\sigma_\A}{\sigma_\D}$
by HERMES using some assumptions$^{\mathrm{F}10}$ 
(in the upper part of their Fig.~5).\cite{ref1}
It is possible that $\frac{\tilde{R}^\A}{\tilde{R}^\D}$
falls again as $Q^2$ approaches zero. Furthermore it is notable that the extracted
increase\cite{ref1} in the ratio in going from He to $^{14}$N is approximately given
by $\frac{\delta(A=14)}{\delta(A=4)}\cong 2$, in accord with Eq.~(5).
The interpretation of the data in terms
of $\tilde{R}^\A > \tilde{R}^\D$ is strengthened by the likelihood of an enhancement in
the interaction of longitudinal photons with the correlated sea among three nucleons,
for the reasons given below. With respect to the $Q^2$ domain of the HERMES
data at small $\langle x\rangle$, one might assume that the multiplicative factor
in Eq.~(3) is unity, and attribute the behavior of their basic data for 
$\frac{\sigma_\A(x,Q^2)}{\sigma_\D(x,Q^2)}$ shown here in Figs.~1,2, to an unusual, strong variation
with low $Q^2$, at small $x$, of $\frac{F_2^\A(x,Q^2)}{F_2^\D(x,Q^2)}$, as in a form
factor with the general limiting behavior of $F(Q^2)$ in Eq.~(2). However,
then the $Q^2$ dependence of the data taken at higher beam energies\cite{ref4,ref5} is
not represented at small $\langle x\rangle$ and moderate $Q^2$ (by our simple $F(Q^2)$).
\par
In fact, assuming the validity of the interpretation in terms of 
$\tilde{R}^\A > \tilde{R}^\D$, also allows for a definite argument that there is new physics
involving multinucleon correlations from explicit quark-gluon dynamics in nuclei.
Interaction of the virtual photon with a single ``isolated'' quark, largely involves
no helicity-flip; then angular momentum conservation requires that the interaction
be with a transverse photon (as is most easily seen in the ``brick wall'' system
for the photon-quark collision). This constraint does not hold in a multiquark
correlation, where the struck quark is not ``free'', but rather propagates, emits
a gluon, and propagates to another nucleon. An exercise involving  only 
Clebsch-Gordon coefficients indicates that if the spin-orientation of an interacting
system of three quarks (nucleons) does not change, then interaction with a longitudinal
photon (behaving like a spin-zero quanta) is favored over a transverse photon by
an {\it a priori} factor of $\frac{3}{2}$. There can also be a dynamical enhancement
of longitudinal photons in interaction with aggregates of sea. By comparing initial-interaction
coefficients of longitudinal and transverse virtual-photon polarization, we estimate
an enhancement factor of the order of 
\begin{displaymath}
\left\{ \frac{\left((m_N x)^2 + \tilde{m}_q^2\right)^{1/2}}{(m_Nx)}  
\tilde{f}(Q^2)\right\}^2 \sim 2,
\end{displaymath}
in the small-$\langle x\rangle$, low-$Q^2$ domain of HERMES, for an effective
aggregate mass $\tilde{m}_q$ of $\sim 20$ MeV ($\tilde{f}$ is a form factor), 
with $m_N$ the nucleon mass (therefore $m_Nx\sim 20$ MeV for $x\sim 0.02$).
\par
A final remark concerns the shadowing at $Q^2=0$ for real photons, when compared
to that for virtual photons with $Q^2$ near to zero. If the explicit partonic
effect first gives increasing shadowing as $Q^2$ moves away from zero, then
this effect on top of the conventional geometric effect already present for real photons,
would give somewhat more shadowing for the virtual photons. There may be some
experimental evidence for this\cite{ref9,ref26}. In addition, since the
geometric shadowing effect at a given very small $x$ is expected to become
less with a marked increase in $Q^2$, if the partonic shadowing effect initially
increases away from $Q^2=0$, then the overall shadowing effect will show 
a reduced variation with increasing $Q^2$.\par
In summary, when viewed and quantitatively correlated within a unified physical
picture, the new data, and all of the earlier improved data from experiments
over the last ten years concerning the EMC effect and shadowing, provide the
best evidence yet for an explicit dynamical role played by exchanged quarks
and nuclear gluons in atomic nuclei.\par
We thank the HERMES collaboration for information, in particular Prof.~M.~D\"uren
and Dr.~G.~van der Steenhoven.

%
%
\newpage
\begin{figure}[t]
\begin{center}
\mbox{\epsfysize 13cm \epsffile{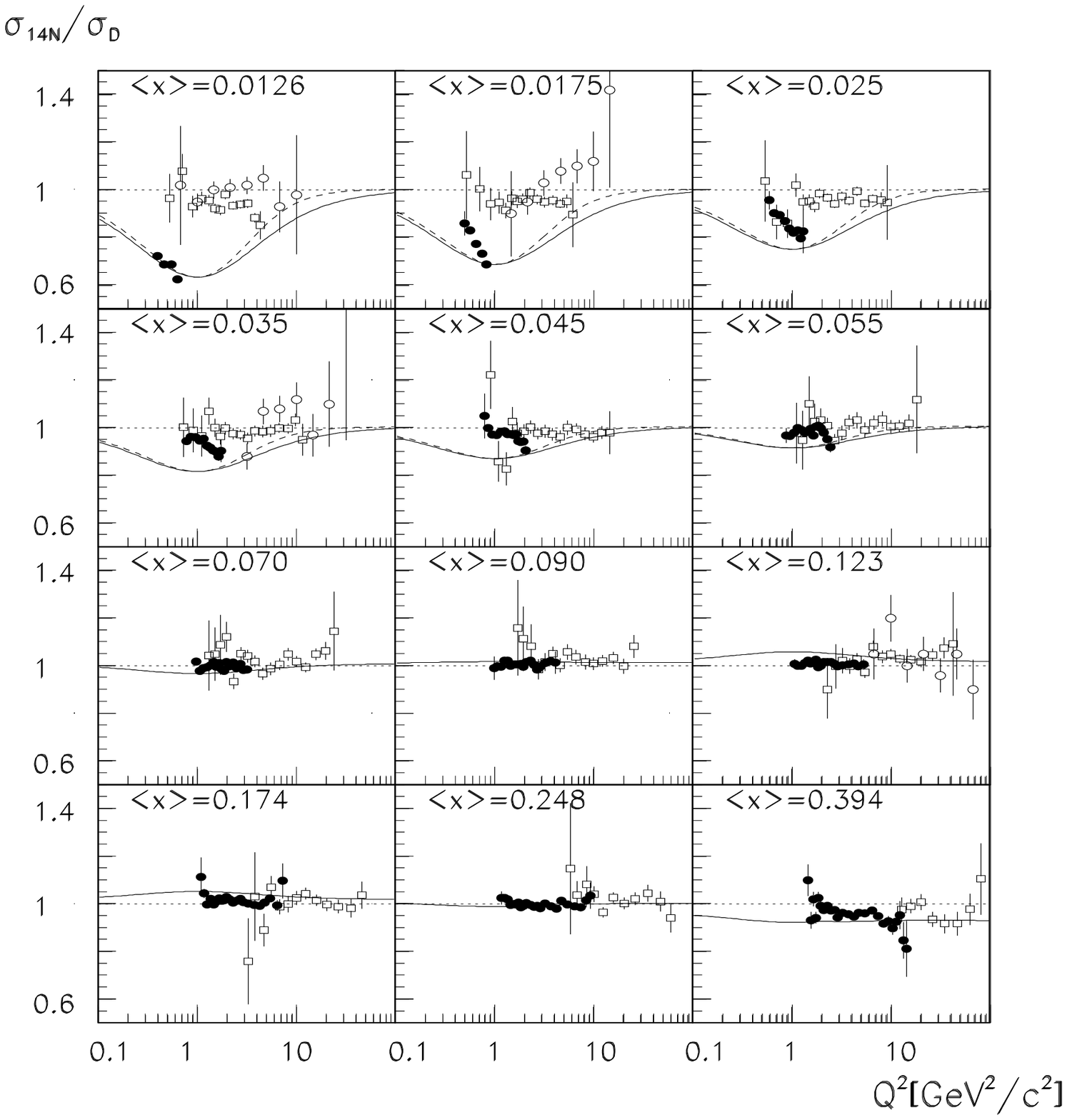}}
\caption{ The ratio of cross section for inclusive deep-inelastic
lepton scattering from $^{14}$N to that of $^2$H, versus $Q^2$ for
specific $x$-bins: the solid circles are HERMES data\protect\cite{ref1}.
The open squares are NMC data\protect\cite{ref4} for $^{12}$C and the 
open circles are E665 data\protect\cite{ref5} for $^{12}$C. The solid curves
are calculated from Eqs.~(2,3) for $\A=14$. The dashed curves illustrate
a form factor with a stronger fall-off at large $Q^2$: 
$F(Q^2)=\frac{27Q^2}{(Q^2+2)^3}$.}
\end{center}
\end{figure}
%
%
\begin{figure}[t]
\begin{center}
\mbox{\epsfysize 13cm \epsffile{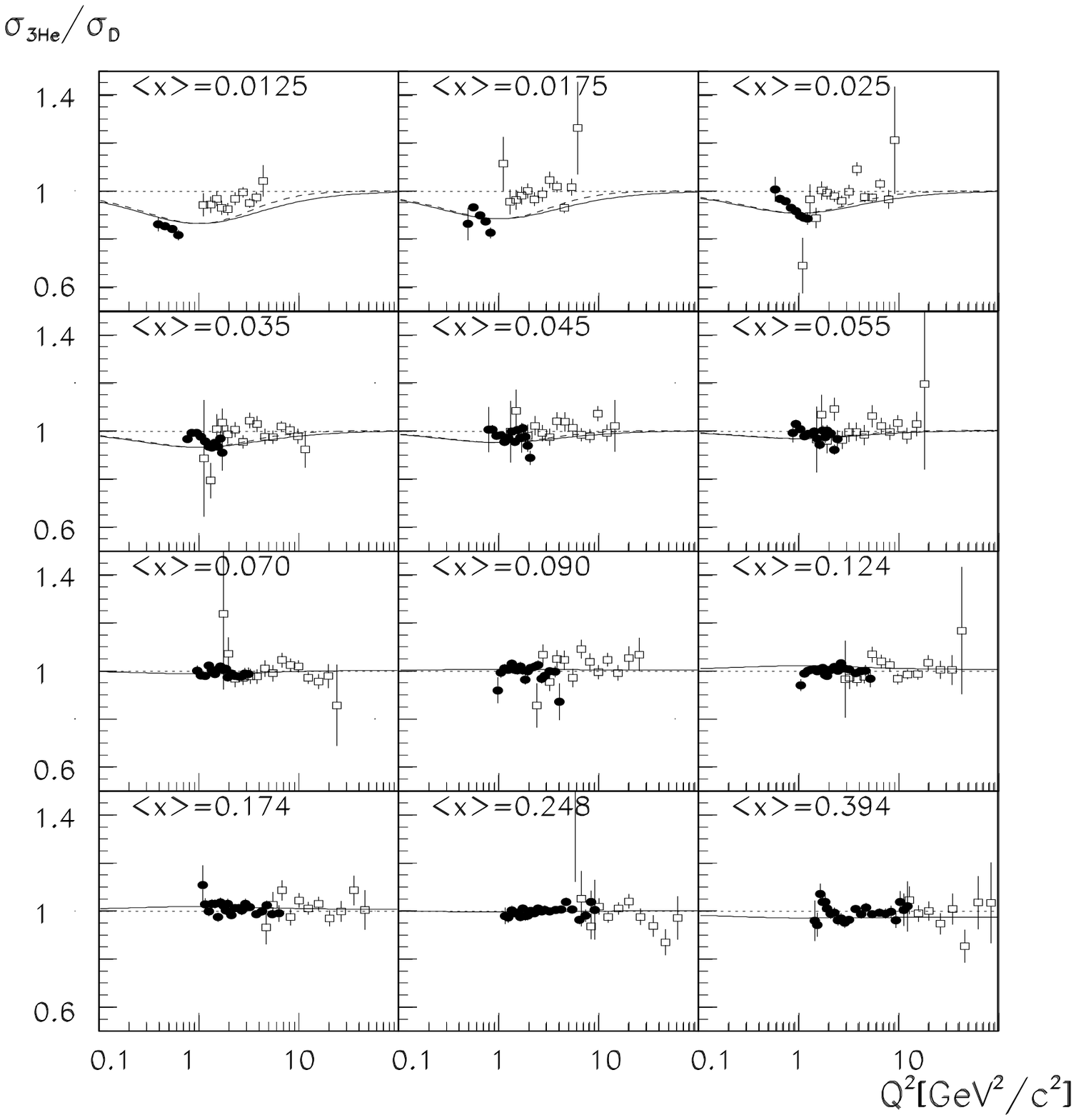}}
\caption{ The ratio of cross section for inclusive deep-inelastic
lepton scattering from $^3$He to that for $^2$H, versus $Q^2$ for
specific $x$-bins: the solid circles are preliminary HERMES data  
(http://dxhra1.desy.de/notes/pub/trans-public-subject.html).
The open squares are NMC data\protect\cite{ref4} for $^4$He. The solid curves
are calculated from Eqs.~(2,3) for $\A=3$. The dashed curves illustrate
a form factor with a stronger fall-off at large $Q^2$: 
$F(Q^2)=\frac{27Q^2}{(Q^2+2)^3}$.}
\end{center}
\end{figure}

\begin{thebibliography}{99}
\bibitem{ref1}HERMES Collab, K.~Ackerstaff et al., hep-ex/9910071, Nov.~1999
\bibitem{ref2}J.~J.~Aubert et al.~(EMC), Phys.~Lett.~{\bf B123} (1983), 275
\bibitem{ref3}A.~Bodek et al., Phys.~Rev.~Lett.~{\bf 50} (1983), 1431; {\bf 51} (1983), 534\\
R.~G.~Arnold et al.~, Phys.~Rev.~Lett.~{\bf 52} (1984), 727
\bibitem{ref4}P.~Amaudruz et al.~(NMC), Nucl.~Phys.~{\bf B441} (1995) , 3
\bibitem{ref5}M.~R.~Adams et al.~(E665), Z.~Phys.~{\bf C67} (1995), 403
\bibitem{ref6}J.~Gomez et al.~(SLAC), Phys.~Rev.~{\bf D49} (1994), 4348
\bibitem{ref8}This is emphasized, with a summary of the relevant data, in S.~Barshay
and D.~Rein, Particle World {\bf 4} (1994), 3
\bibitem{ref9} J.~Franz et al., Z.~Phys.~{\bf C10} (1981), 105\\
J.~Bailey et al., Nucl.~Phys.~{\bf B151} (1979), 367\\
J.~Eickmeyer et al., Phys.~Rev.~Lett.~{\bf 36} (1976), 289\\
S.~Stein et al., Phys.~Rev.~{\bf D12} (1975), 1884
\bibitem{ref10} M.~S.~Goodman et al., Phys.~Rev.~Lett.~{\bf 47} (1981), 293
\bibitem{ref11}S.~Barshay and D.~Rein, Z.~Phys.~{\bf C46} (1990), 215
\bibitem{ref12}S.~Barshay, Phys.~Lett.~{\bf 100B} (1981), 276
\bibitem{ref14}G.~Smirnov, Phys.~Lett.~{\bf B364} (1995), 87\\
G.~Smirnov, Yad.~Fiz.~{\bf 58}, no.~9 (1995), 1712; Phys.~At.~Nucl.
(Engl.~translation) {\bf 58}, no.~9 (1995)\\
See also hep-ph/9611203. This author has also argued from the data for a
basic dynamical role in the He nuclei.
\bibitem{ref15} P.~Amaudruz, et al.~(NA37/NMC), Z.~Phys.~{\bf C53} (1992), 73
\bibitem{ref17} CDHS Collab.~, J.~G.~H.~de Groot et al., Z.~Phys.~{\bf 1} (1979), 143\\
H.~Abramowitz et al., Z.~Phys.~{\bf C17} (1983), 283
\bibitem{ref26}L.~Criegee et al., Nucl.~Phys.~{\bf B121} (1977), 38
\end{thebibliography}
\end{document}